\documentclass[onecolumn]{emulateapj}
\pdfoutput=1
\usepackage{graphics}

\newcommand{\avg}[1]    {{\langle #1 \rangle}} 
\newcommand{\beq}	{\begin{equation}}
\newcommand{\eeq}	{\end{equation}}
\newcommand{\beqa}	{\begin{eqnarray}}
\newcommand{\eeqa}	{\end{eqnarray}}
\newcommand{\dis}{\displaystyle}
\newcommand{\calm}	{{\cal M}}
\newcommand{\mrms}      {{\calm_{\rm rms}}}

\newcommand{\vecB}	{{\bf B}}
\newcommand{\vecF}	{{\bf F}}

\newcommand{\vecv}	{{\bf v}}

\newcommand{\curl}      {{\bf\nabla\times}}
\newcommand{\vectimes}	{{\bf \times}}

\newcommand{\gad}       {{\gamma_{\rm AD}}}

\newcommand{\Gad}       {{\Gamma_{\rm AD}}}
\newcommand{\Gt}        {{\Gamma_t}}

\newcommand{\alfven}    {{Alfv$\acute{\rm e}$n }}

\newcommand{\alfvenic}  {{Alfv$\acute{\rm e}$nic }}

\newcommand{\chio}	{\chi_{i0}}

\newcommand{\nh}	{n_{\rm H}}
\newcommand{\rad}	{R_{\rm AD}}

\newcommand{\rado}      {R_{\rm AD,\, 0}}

\newcommand{\tad}	{t_{\rm AD}}

\newcommand{\tf}	{t_f}

\newcommand{\va}	{v_{\rm A}}

\newcommand{\vrms}	{v_{\rm rms}}

\newcommand\brms        {B_{\rm rms}}

\newcommand{\e}	        {$^{-1}$}
\newcommand{\ee}	{$^{-2}$}
\newcommand{\eee}	{$^{-3}$}

\newcommand{\ld}	{\ell_{d}}
\newcommand{\ldb}	{\ell_{\delta B}}

\newcommand{\ma}	{{\calm_{\rm A}}}

\newcommand{\nbh}	{\bar n_{\rm H}}

\newcommand{\radl}	{R_{\rm AD}(\ell)}

\newcommand{\epsad}		{\epsilon_{\rm AD}}
\newcommand{\epst}		{\epsilon_t}
\newcommand{\radlo}     {R_{\rm AD}(\ell_0)}
\newcommand{\radld}     {R_{\rm AD}(\ell_d)}
\newcommand{\radolo}     {R_{\rm AD,\,0}(\ell_0)}

\newcommand{\ldi}	{\ell_{\delta B,\,\infty}}
\newcommand{\lint}	{L_{\rm int}}
\newcommand{\ldpc}	{\ell_{\rm d,\,pc}}

\newcommand{\vda}	{\avg{v_d^2}}

\shorttitle{Ambipolar Diffusion Heating in Turbulent Systems}
\shortauthors{Li, Myers, AND McKee}
\begin{document}
\title{AMBIPOLAR DIFFUSION HEATING IN TURBULENT SYSTEMS}
\author{Pak Shing Li}
\affil{Astronomy Department, University of California,
    Berkeley, CA 94720}
\email{psli@astron.berkeley.edu}
\author{Andrew Myers}
\affil{Physics Department, University of California, Berkeley, CA 94720}
\email{atmyers@berkeley.edu}
\and
\author{Christopher F. McKee}
\affil{Physics Department and Astronomy Department, University of California,
    Berkeley, CA 94720}
\email{cmckee@berkeley.edu}

\begin{abstract}
The temperature of the gas in molecular clouds is a key determinant of the characteristic mass of star formation.  Ambipolar diffusion (AD) is considered one of the most important heating mechanisms in weakly ionized molecular clouds.  In this work, we study the AD heating rate using 2-fluid turbulence simulations and compare it with the overall heating rate due to turbulent dissipation.  We find that for observed molecular clouds, which typically have \alfven Mach numbers of $\sim 1$ \citep{cru99} and AD Reynolds numbers of $\sim 20$ \citep{mck10}, about 70\% of the total turbulent dissipation is in the form of AD heating.  AD has an important effect on the length scale where energy is dissipated: when AD heating is strong, most of the energy in the cascade is removed by ion-neutral drift, with a comparatively small amount of energy making it down to small scales. We derive a relation for the AD heating rate that describes the results of our simulations to within a factor of two. Turbulent dissipation, including AD heating, is generally less important that cosmic-ray heating in molecular clouds, although there is substantial scatter in both. 
\end{abstract}

\section{INTRODUCTION}

The temperature of the gas in molecular clouds is readily observable and is a key determinant
of the characteristic mass of star formation, since the Jeans mass varies as the 3/2 power of the
temperature. 
In molecular gas, which is generally shielded from far-ultraviolet radiation, the heating
processes that determine the temperature are cosmic-ray ionization, turbulent dissipation,
ambipolar diffusion, magnetic reconnection, and hydrodynamic compression. Cosmic rays are generally assumed to
be the dominant heating mechanism (e.g., \citealp{gol01}), but \citet{pan09} found that
turbulent dissipation can give a higher heating rate in some cases. Ambipolar diffusion (AD) is
the slippage of magnetized ions through the dominant neutral gas, and 
the collisions resulting from this relative motion generate heat. AD heating is considered to be one of the most important heating mechanisms in molecular clouds \citep[e.g.,][]{sca77,zwe83,elm85,pad00}. \citet{pad00} (hereafter PZN00) studied the AD heating rate in simulations of turbulence in a periodic box and found that the AD heating rate was significant.

Although AD heating and turbulent dissipation have generally been considered as separate processes,
in a turbulent medium AD heating is a component of turbulent dissipation. Turbulence is driven
on large scales and cascades down to smaller scales. This picture is rigorously true for incompressible
turbulence and has been shown to be consistent with numerical simulations for supersonic
(compressible) turbulence \citep{kri07}. AD occurs on the scale of the neutral-ion mean free path, which for
weakly ionized gas is much greater than the neutral-neutral mean free path that governs viscous
dissipation. As a result, energy is drained from the turbulent cascade first by AD and then by
viscous dissipation. Magnetic reconnection also contributes to turbulent dissipation on small scales.
 
In view of the importance of AD heating in turbulent media, we have re-evaluated its magnitude with higher-resolution simulations than were possible a decade ago. 
The AD heating rate per unit volume, $\Gad$, due to this ion-neutral friction is
\beq
\Gad = \gad \rho_i \rho_n \left(\vecv_i -\vecv_n \right)^2 = \gad \rho_i \rho_n v_d^2.
\label{eq:adheat}
\eeq
where $\rho_{i,n}$ and $v_{i,n}$ are the density and velocity of the ion and neutral components, respectively, 
$\gad = \langle \sigma v \rangle / (m_n + m_i)$ is the ion-neutral coupling constant, and $v_d$ is the magnitude of the ion-neutral
drift velocity.
The parameter $\langle \sigma v \rangle$ is the collision rate coefficient between ionic and neutral species, where $m_n$ and $m_i$ are the mean neutral and ion masses, respectively.
We are interested in the case in which the ion mass fraction, $\chi_i=\rho_i/\rho_n$,  is sufficiently low that the ion inertia is negligible. Solving the full two-fluid MHD equations is computationally prohibitive in this case, since the ion \alfven velocity $\va\propto \chi_i^{-1/2}$ is very large and the Courant limit on the time step correspondingly very small. Two approximations have been used to treat this problem:
one is to use a single-fluid approximation that turns the induction equation into a diffusion equation; the
difficulty with this approach is that the time step scales as the square of the grid spacing, and becomes
prohibitively small at high resolution. The other approach is to solve the full 2-fluid MHD equations with a semi-implicit approach that uses the heavy-ion approximation \citep{li06, ois06}, in which the ion mass fraction, $\chi_i$, is raised and the coupling coefficient, $\gad$, is lowered in a way that preserves their product, leaving the ion-neutral drag unchanged. We have used this approach to perform high resolution non-ideal MHD turbulence simulations (\citealp{li08} (Paper I), \citealp{mck10} (Paper II), \citealp{li12} (Paper III)). It allows us to use explicit time stepping for both fluids without running afoul of the CFL constraint due to high ion \alfven speeds. \citet{li06} give a justification of this approach as well as a more detailed description of the equations and numerical techniques employed.

In this paper, we evaluate the importance of AD heating and compare it with the overall turbulent dissipation rate. In Section 2, we summarize our numerical models and the assumptions adopted.  In Section 3, we report our simulation results and demonstrate convergence.  In Section 4, we provide an analytic model that predicts the mean AD heating rate and show that it is consistent with the numerical results. This model gives the approximate heating rate in terms of observable molecular cloud properties only, so that the importance of AD drift heating can be gauged for real molecular clouds. 
In Section 5 we discuss our results and compare our prediction on the AD heating rate to the work of PZN00, as updated in PZN12 \citep{pad12}. Section 6 summarizes our conclusions.

\section{NUMERICAL MODELS AND CONVERGENCE STUDY}

The results discussed in this paper are based on the series of $512^3$ AD, ideal MHD, and hydrodynamic turbulent box simulations investigated in Papers II and III, plus several additional AD runs with different magnetic field strengths and resolutions. In all runs, the gas was isothermal, there was no gravity, and the boundaries were periodic. In the MHD runs, the field was initially uniform. The field strength is characterized by the plasma $\beta$ parameter,
\beq
\beta=\frac{8\pi\bar\rho c_s^2}{\brms^2}=2\;\frac{\ma^2}{\calm^2},
\eeq
where $\bar\rho$ is the mean density in the box, $c_s$ is the isothermal sound speed,
$\ma$ is the \alfven Mach number and $\calm$ is the three-dimensional rms sonic Mach number.
In Papers I-III we did not distinguish the equilibrium value of $\beta$ from the initial value, $\beta_0$
because $\brms$ did not differ significantly from the initial field strength, $B_0$, but here we do.

To characterize the importance of AD in these simulations, we use the AD Reynolds number, which is the ratio of the characteristic AD timescale to the flow time, or, equivalently, the ratio of the size of the system to the characteristic AD length scale \citep{zwe97},
\beq
\radlo\equiv\frac{4\pi\gad\bar\rho_i\bar\rho_n \ell_0 \vrms}{\brms^2} = \frac{\tad}{\tf}.
\label{eq:radell}
\eeq
Here, $\bar\rho_{i,\, n}$ is the mean density of the ions and neutrals, respectively, $\ell_0$ is the size of the system (for the simulation, it is the size of the turbulent box), 
$\brms$ is the rms magnetic field, and
\beq
\vrms^2\equiv\frac{1}{\bar\rho}\int \rho v^2 dV
\eeq
is the density-weighted mean-squared velocity of the system.
The dynamical crossing time is $\tf=\ell_0/\vrms$, and $\tad$ is the AD time scale. We denote the initial value of the AD Reynolds number by $\radolo$.

The turbulence in all of the models was maintained at a constant rms Mach number, $\calm$, by a fixed driving pattern generated using the recipe of \citet{mac99}.  The turbulence was driven between wavenumbers $k = 1 \sim 2$ 
(all wavenumbers are in units of  $2\pi/\ell_0$) for a period of 3$\tf$. We allowed the turbulence to develop for 1$\tf$, and averaged all results over 14 data dumps spread over the subsequent 2$\tf$ unless otherwise indicated.
Both the neutral and ionized components of the gas were driven using the same driving pattern and amplitude in order to prevent the driving from creating an artificial velocity difference between the two components.
In the ideal run and five main AD runs, the turbulent box was initially threaded by a uniform magnetic field 
with $\beta_0 = 0.1$.

In addition to the five main sub-\alfvenic AD models, we carried out three additional AD simulations.  Two models had $\calm = 3$ but $\beta_0=1$ and 10 to provide information on slightly and highly super-\alfvenic turbulence.  A third AD model driven to $\calm = 10$ at $256^3$ resolution provides a high thermal Mach number model for comparison. Table 1 summarizes the parameters of all the runs in this paper.

\begin{table}
\begin{center}
\caption{Model Parameters \label{tbl-1}}
\begin{tabular}{llllllclcc}
\\
\tableline\tableline
Model$^{\rm a}$ & $\gad$ & $\rado(\ell_0)$ & $\rad(\ell_0)$ & $\beta_0$ & $\beta$ & $\brms/B_0$ & $\ma$ & $\lint/\ell_0$ & $resolution^{\rm b}$\\
\tableline
m3ph     &-        &0        &0        &0    &0     &-    &0     & 0.70 & $512^3$\\
m3c2r-1  &4        &0.12     &0.12     &0.1  &0.1   &1.00 & 0.67 & 0.71 & $512^3$\\
m3c2r0   &40       &1.2      &1.2      &0.1  &0.1   &1.00 & 0.67 & 0.73 & $512^3$\\
m3c2r1   &400      &12       &11.6     &0.1  &0.097 &1.02 & 0.66 & 0.88 & $512^3$\\
m3c2r2   &4000     &120      &113    &0.1  &0.096 &1.03 & 0.66 & 0.96 & $512^3$\\
m3c2r3   &40000    &1200     &1110   &0.1  &0.095 &1.03 & 0.65 & 0.92 & $512^3$\\
m3i  	 &$\infty$ &$\infty$ &$\infty$ &0.1  &0.092 &1.04 & 0.65 & 0.87 & $512^3$\\
m3c2r1b0 &40       &12       &8.7      &1.0  &0.75 &1.15 & 1.84 & 0.86 & $512^3$\\
m3c2r1b1 &4        &12       &6.5      &10   &5.4 &1.36 & 5.07 & 0.99 & $512^3$\\
m3c2r2b1 &40       &120      &22     &10   &1.84 &2.33 & 3.07 & 1.01 & $512^3$\\
m10c2    &40       &4        &3.7      &0.1  &0.092 &1.04 & 2.14 & 0.77 & $256^3$\\
\tableline
\end{tabular}
\end{center}
$^{\rm a}$ Models are labeled as ``mxcyrn," where $x$ is the thermal Mach number, $y=|\log\chio|$, and $n=\log(\radl/1.2)$. Model ``m3i" is an ideal MHD and ``m3ph" is a pure hydrodynamic model. Model m3c2r0 is the same as model m3c2h in LMKF. Models m3c2r1b1 and m3c2r1b2 have different plasma $\beta_0$ from the other models. \\
$^{\rm b}$ Except model m10c2, all AD models have $128^3$ and $256^3$ resolution runs for convergence study.
\end{table}

The AD models are based on the assumption that the ions are a separately conserved fluid. A discussion of the effect of assuming ionization equilibrium instead can be found in the Appendix of Paper I and in Paper II. For our AD models, there is no major difference between ion conservation and ionization equilibrium for most of the turbulence statistics, except for the clear differences in the ion density PDF.  With ionization equilibrium, the velocity power spectra are about 1 $\sigma$ steeper than for the ion conservation case. The differences in the properties of the clumps investigated in the above papers between the two ionization models were at the few percent level.  Comparison of the AD heating rates with ion conservation vs. ionization equilibrium for our five main AD models, all at $256^3$ resolution shows that the differences are $\sim 10-15 \%$, which is about 1 $\sigma$.  

\begin{figure}
\begin{center}
\epsscale{.80}
\includegraphics[scale=0.4]{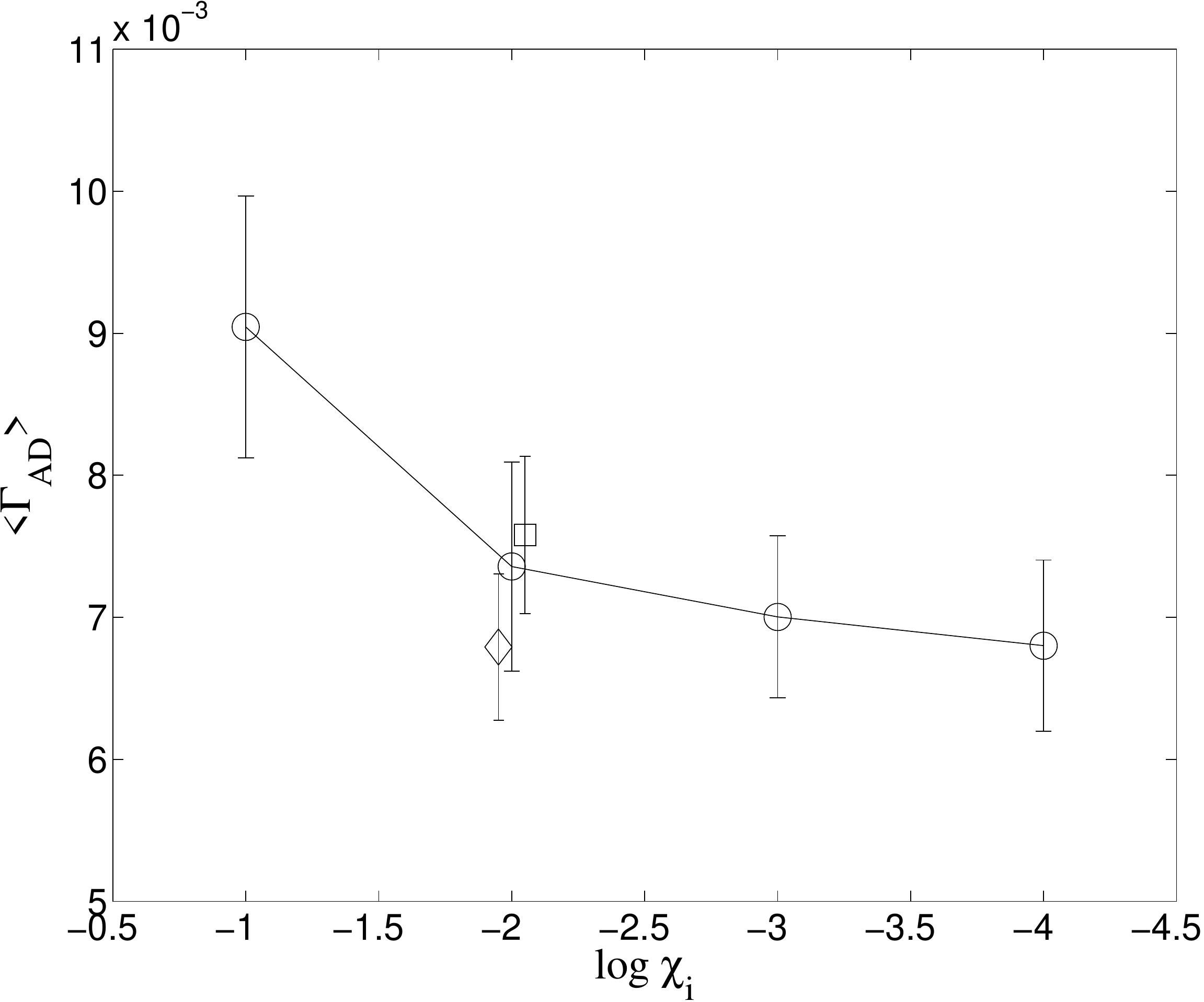}
\caption{Convergence study of the volume mean AD heating rate, $\langle\Gad\rangle$, on resolution (diamonds for $128^3$, circles for $256^3$, and squares for $512^3$) and $\chi_i$.  Using $\chi_i = 0.01$ is accurate enough for the heavy-ion approximation when computing $\langle\Gad\rangle$, and $512^3$ resolution appears to be well-converged.  The data point from the $512^3$ model is slightly shifted to the right and the $128^3$ model to the left of $\chi_i = 0.01$ to prevent the error bars from overlapping.
\label{fig1}}
\end{center}
\end{figure}

We performed a convergence study of the time- and volume-averaged AD heating rate, $\langle\Gad\rangle$, in both spatial resolution and ion-mass fraction ($\chi_i$) to ensure that the results are spatially resolved and that the heavy-ion approximation is accurate for our chosen value of $\chi_i = 0.01$.  
To test the heavy-ion approximation, we used the four $256^3$ models, m3c1r0
(corresponding to $\calm=3$, $\chi_i=10^{-1}$ and $\radolo=1.2\times 10^0$)
to m3c4r0 ($\chi_i=10^{-4}$), from Paper I and plot $\langle\Gad\rangle$ vs. $\chi_i$ in Figure \ref{fig1}.  
(Keep in mind that the physical values of $\chi_i$ can be less than $10^{-6}$.) The results show that that using $\chi_i = 0.01$ in the heavy-ion approximation gives a value of $\Gad$ that is accurate to within 10\% in this case.
We have also run $128^3$ and $256^3$ simulations for the five main AD models to study the convergence behavior of $\langle\Gad\rangle$ with spatial resolution.
The values of $\langle\Gad\rangle$ for the $128^3$ and the $512^3$ simulations of model m3c2r0 are shown in Figure \ref{fig1} for illustration.  The heating rate shows the expected second-order convergence, within the uncertainties, and we conclude that a numerical resolution of $512^3$ provides a well-converged AD heating rate.  We obtained similar results for all but the $\radolo = 1200$ case, which showed convergence at an order of only $\sim$ 1.5.  We will use Richardson extrapolation \citep{ric27} to estimate the converged values of the AD heating rate in the modeling in Section \ref{sec:modeling}.

\section{RESULTS FOR THE TURBULENT DISSIPATION RATE AND THE AD HEATING RATE}
\label{sec:results}

Our main objectives are to determine the total time-averaged rate of dissipation of turbulent energy, $\avg{\Gt}$, and that part of it that is due to ambipolar diffusion, $\avg{\Gad}$, as a function of the AD Reynolds number, $\radlo$. We begin with the total time-averaged rate of dissipation of turbulent energy, which can be written as
\beq
\avg{\Gt} = \epsilon_t \;\frac{\bar\rho \vrms^3}{\ld},
\label{eq:gt}
\eeq
where $\ld$ is the typical outer length scale of the turbulence;
in numerical simulations, this is the typical length scale on which the turbulence is
driven, which in our case is $\ell_0/\sqrt{2}$ ($k = 1 \sim 2$).
In studies of incompressible turbulence, the dissipation rate is often normalized to the integral
length scale, $\lint$. We have chosen to normalize to the typical driving scale, $\ld$, as has
been done in some past simulations of supersonic turbulence \citep[e.g.,][]{mac99,lem09}
From our pure HD, ideal MHD, and all AD turbulence models at $512^3$, we find that
$\avg{\lint}/\ell_0 = 0.85\pm0.15$.
We have not bothered to distinguish $\vrms$ from its time average, $\avg{\vrms}$, since the turbulent driving forces them to be equal to within 0.01\%

First, we discuss the turbulent dissipation rate in the HD and ideal MHD cases.
Our HD model gives $\epst\simeq 0.65$, which is consistent with the results of other turbulence simulations at similar resolution \citep[e.g.,][]{mac99,lem09}. 
We have not carried out a convergence study of our hydrodynamic result, but we note that the dissipation
rate for this simulation agrees with that for the low $\radlo$ simulation m3c2r-1, which we verified is converged.
The dissipation rate for incompressible HD turbulence is smaller:
\citet{kan03} carried out simulations with far higher resolution (up to $4096^3$) for this case and found $\epst\simeq 0.4-0.5$.
They used the integral length scale in computing the dissipation rate; for our pure HD model,
this is almost the same as $\ld$.
Our ideal MHD model gives $\epst\simeq 0.25$, which is at the low end of Mac
Low's (1999) results ($\epst\sim 0.3-0.6$) from his lower resolution 
simulations of supersonic MHD turbulence.  In their simulations of such turbulence,  
\citet{lem09} found $\epsilon_t\simeq 0.4-0.5$, about 50\% larger than the value we find. This is 
probably because of the difference in the driving method:  \citet{lem09} used a variable, sharply peaked driving pattern,
while we used a fixed, flat-top driving pattern.
They measured the dissipation time in terms of the flow time across $\lambda_{\rm pk}$, the 
wavelength of the peak in 
the perturbation spectrum; this is equivalent to replacing $\ld$ in Equation (\ref{eq:gt}) by $\lambda_{\rm pk}$.
They found that the normalized dissipation time was very insensitive to changes in $\lambda_{\rm pk}$.
Studies of the energy dissipation rate in simulations of supersonic driven turbulence with and without magnetic fields have found that the energy dissipation rate for ideal MHD turbulence is smaller than that for HD turbulence by a factor of $1.4 \sim 2$ \citep[e.g.,][]{mac99,lem09}, and our result is at the upper end of this range.

The total dissipation rate is equal to the sum of the AD dissipation rate, $\Gad$, and the other forms of dissipation, which we shall group together in $\Gamma_{\rm other}$. In the simulations, $\Gamma_{\rm other}$ is a combination of numerical viscosity and numerical resistivity that arise from our imperfect discretization of the fluid equations. In real partially ionized, turbulent fluids such as molecular clouds, $\Gamma_{\rm other}$ represents energy that cascades down to very small scales and is dissipated by physical processes such as molecular viscosity and Ohmic dissipation.  Our convergence study shows that whatever the details of the numerical dissipation, it occurs on sufficiently small scales in the simulation that it has a small effect on the mean AD heating rate. 

\begin{figure}
\begin{center}
\epsscale{.80}
\includegraphics[scale=0.4]{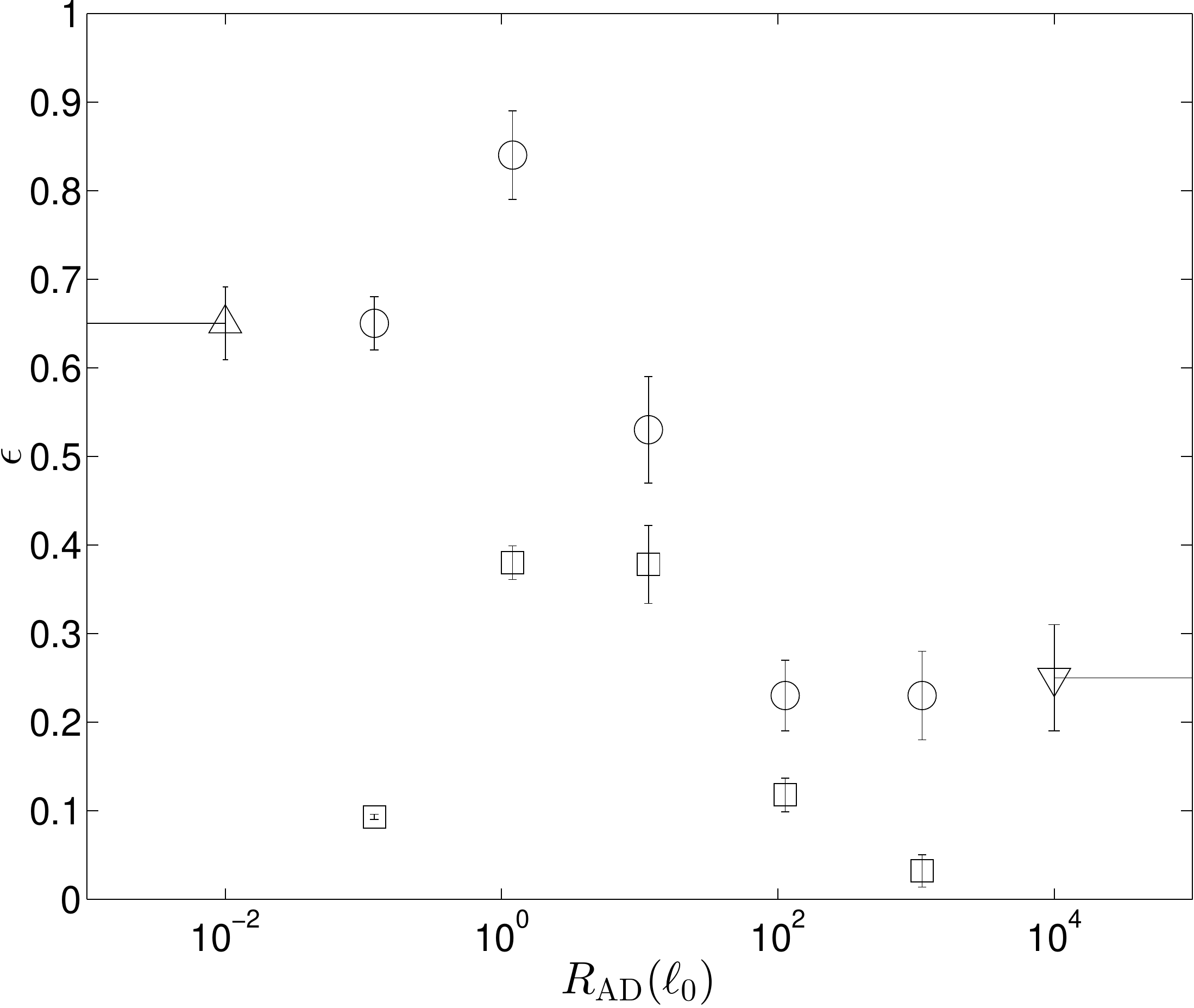}
\caption{Normalized AD heating rates (squares) and total energy injection rates
(circles) of the five AD models 
with $\beta_0 = 0.1$ as a function of $\radlo$. The energy injection rates of an ideal MHD (down triangle) model and a pure hydrodynamics (up triangle) model are also plotted for comparison.  The $\radlo$ of these two limiting cases are moved from $\infty$ to $10^4$ for the ideal MHD model and from 0 to $10^{-2}$ for the pure hydro model to fit on the plot.  The total energy injection rates of the two limiting cases match the AD models with largest and smallest $\rad$.  
The drop in AD heating rate in the model with $\radlo\simeq 0.1$ (m3c2r-1) is due to the saturation of the drag velocity $v_d$ in the weakly coupled ion and neutral components; the drop in the models with $\radlo\simeq 10^2,\, 10^3$ is due to the reduction in $v_d$ when the ions and neutrals are well-coupled.   See Section \ref{sec:results} for discussion.
\label{fig2}}
\end{center}
\end{figure}

To facilitate comparison with the total dissipation rate, we normalize 
the time-averaged AD heating rate, $\avg{\Gamma_{\rm AD}}$, by $\rho \vrms^3/\ld$, so that
\beq
\avg{\Gad} = \epsilon_{\rm AD} \;\frac{\bar\rho \vrms^3}{\ld}.
\eeq
We compute $\epsilon_{\rm AD}$ for each of the five AD models with $\beta_0 = 0.1$ and plot the results versus $\radlo$ in Figure \ref{fig2}.
We also plot the normalized total energy injection rate, $\epsilon_{\rm t}$, 
based on the time-averaged energy driven into the box to maintain the gas at Mach 3.
The uncertainty in $\epsilon$ is given by the standard error of the mean, which we calculate as the standard deviation evaluated for a total 14 data sets in the last $2\tf$ divided by the square root of the number of independent samples of $\epsilon$,
which we estimate as 2.

\begin{figure}
\begin{center}
\epsscale{.80}
\includegraphics[scale=0.6]{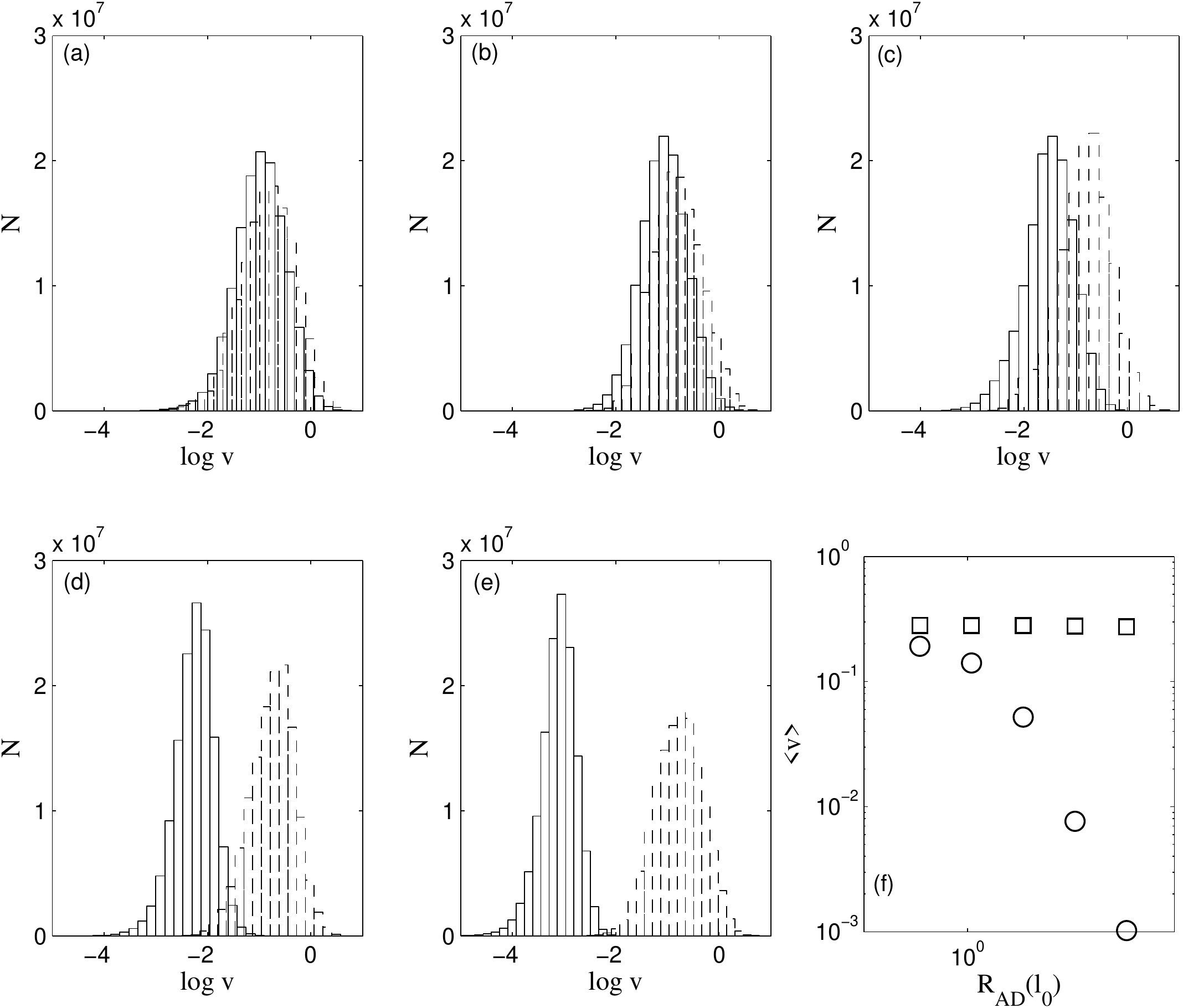}
\caption{Distributions of the magnitudes of the drag velocity,
 $|v_d|$ (solid line) and the neutral velocity, $|v_n|$ (dashed line), of models (a) m3c2r-1 ($\radolo=0.12$), (b) m3c2r0 ($\radolo=1.2$), (c) m3c2r1 ($\radolo=12$), (d) m3c2r2 ($\radolo=120$), and (e) m3c2r3 ($\radolo=1200$).  
(f) The normalized dispersion in the drift velocity (circles), $v_{d\,\rm rms}/c_s$, and in the neutral velocity (squares),
$v_{n\,\rm rms}/c_s\simeq \calm$ as functions of $\radlo$.}
\label{fig3}
\end{center}
\end{figure}

For the five AD models, $\epsilon_{\rm t}$ increases with decreasing $\radlo$ from 0.25 (the same as the ideal MHD value) at the highest value of $\radlo$ to 0.84 when $\radlo \sim 1$; it then drops back to about 0.65  (the same as the HD value) as $\radlo \rightarrow 0$. The increase in $\epsilon_{\rm t}$ at intermediate values of $\radlo$ is clearly due to the additional energy lost to AD heating.  The AD heating is small in the model with the highest value of $\radlo$ ($\simeq  1000$), which is expected as ions and neutrals are strongly coupled.  When $\radlo$ becomes lower and the drift velocity larger, $\langle\Gad\rangle$ increases, reaching a maximum around the models with 
$\radlo = 11.6$ and $1.2$.
At yet lower values of $\radlo$, the drag velocity saturates at the rms velocity and the ion-neutral collision
rate declines (in molecular clouds, this would most likely be due to a decrease in the ionization), resulting
in a decreasing value of $\epst$. Figure \ref{fig3} illustrates the saturation of the density-weighted $|v_d|$.  As $\radlo$ becomes small, the distribution of $|v_d|$ almost overlaps the neutral velocity distribution. The volume-weighted velocity distributions also show the same trend of saturation of the magnitude of the drift velocity, $|v_d|$.  Figure \ref{fig3}(f) shows the density-weighted $v_{d,{\rm rms}}$ and $v_{n,{\rm rms}}$ of the five AD models.  

\section{MODELING THE AD HEATING RATE}
\label{sec:modeling}

Here we develop an analytic model for the AD heating rate, which depends on the ion-neutral drift
velocity $\vecv_d=\vecv_i-\vecv_n$.
The drag force on the neutrals is $\vecF_{\rm drag}=\gad\rho_i\rho_n\vecv_d$, where $\gad$ is the ion-neutral
coupling coefficient. The rate at which heat is generated by the drag is $\vecF_{\rm drag}\cdot\vecv_d$, so that
the volume-averaged AD heating rate is (Equation (1))
\beqa
\avg{\Gad}&=&\left\langle\gad\rho_i\rho_n\vrms^2\left(\frac{v_d^2}{\vrms^2}\right)\right\rangle,\\
&\sim&\gad\bar\rho_i\bar\rho_n \vrms^2\left(\frac{\avg{v_d^2}}{\vrms^2}\right),
\label{eq:avggad}
\eeqa
where
\beq
\vda=\avg{\rho v_d^2}/\bar\rho
\eeq
is mass-averaged, just like $\vrms^2$ (note that $\avg{v_d^2}$ is the only case in which $\avg{x}$ represents a mass average; otherwise it denotes a volume average). The second step is accurate for small and moderate values of $\radld$, but can be
in error by up to a factor 3 at large $\radld$. We can rewrite this in terms of
$\tau$, the ratio of the flow time at the driving scale, $t_f=\ld/\vrms$,
to the neutral-ion collision time,
\beq
\tau\equiv\frac{t_f}{t_{ni}}=\frac{\radld}{\ma^2},
\eeq
and obtain
\beq
\avg{\Gad}=\tau\left(\frac{\avg{v_d^2}}{\vrms^2}\right)\left(\frac{\bar\rho_n\vrms^3}{\ld}\right).
\eeq

\subsection{ESTIMATE OF $\vda$}

When the ion inertia is negligible, the equation of motion for the ions is
\beq
\gad\rho_i  \rho_n\vecv_d=\vecF_L,
\label{eq:drag}
\eeq
where
\beq
\vecF_L=\frac{1}{4\pi}\left[(\curl\vecB)\vectimes\vecB\right]
\eeq
is the Lorentz force. The mean-squared drag velocity is then
\beq
\avg{v_d^2}=\frac{1}{\bar\rho}\left\langle\frac{\rho |\vecF_L|^2}{(\gad\rho_i\rho_n)^2}\right\rangle
\eeq
from equation (\ref{eq:drag}). 
We divide the magnetic field into a steady component and a variable one,
\beq
\vecB=\avg{\vecB}+\delta\vecB,
\eeq
so that
\beq
\brms^2\equiv\avg{\vecB^2}=B_0^2+\delta\brms^2.
\eeq
Now,
\beq
|\vecF_L|\sim\frac{\brms\delta\brms}{4\pi\ldb},
\eeq
where $\ldb$ is the characteristic scale over which the fluctuating component of the
field varies.  If the heating were wave dissipation, $\ldb$ would be proportional to $k^{-1}$.  We now {\it define} $\ldb$ by inserting this approximate expression into
the expression for the mean-squared drag velocity,
\beq
\vda\equiv\frac{\brms^2\delta\brms^2}{(4\pi\gad\bar\rho_i\bar\rho_n)^2\ldb^2}.
\label{eq:vda4}
\eeq

Letting $\phi_B\equiv \delta\brms/\brms$, we find
\beq
\frac{\vda}{\vrms^2}=\frac{\phi_B^2}{\radld^2}\left(\frac{\ld}{\ldb}\right)^2.
\label{eq:vda3}
\eeq

To evaluate $\phi_B$, we express the energy in the fluctuating field in terms
of the kinetic energy as
\beq
\frac{1}{8\pi}\delta\brms^2=\frac 12 \xi \bar\rho \vrms^2,
\eeq
where the factor $\xi$ measures the deviations from
equipartition (Heitsch et al. 2001; Paper III). Normalizing with respect to the rms field,
we have
\beq
\phi_B^2\equiv \frac{\delta\brms^2}{\brms^2}=\xi\ma^2.
\eeq
At low \alfven Mach numbers, we expect that the energy in the fluctuating
field will be in equipartition with the associated kinetic energy \citep{zwe95}.
For a uniform field, this kinetic energy is in the directions normal to the field,
so $\xi\simeq \frac 23$. 
The velocity field in the simulations is found to be approximately isotropic.
As discussed above, this should remain valid
in the presence of AD provided that the ions and neutrals are well-coupled 
along the field, which they are under the conditions we consider here.
On the other hand, since $\phi_B^2$ cannot exceed unity,
$\xi\simeq 1/\ma^2$ for large values of $\ma$. As a result, we estimate
\beq
\xi\simeq\frac{\frac 23}{1+\frac 23 \ma^2},
\eeq
which agrees with the results of Stone et al. (1998) to within a factor 1.2.
The corresponding result for $\phi_B^2$ is
\beq
\phi_B^2\simeq \frac{\frac 23\ma^2}{1+\frac 23 \ma^2}.
\label{eq:phiB}
\eeq

The final quantity to be evaluated in Equation (\ref{eq:vda3}) is $\ldb$.
In the limit of large $\radld$, $\ldb$ should have a well-defined value, which
we term $\ldi$.
We will make the assumption that $\ldi$ is proportional to the driving scale, with a constant of proportionality called $\alpha$:
\beq
\ldi = \alpha \ld.
\eeq The motivation for this is that in a periodic box, gradients cannot exist on scales larger than $\ell_0 / 2 \pi$, and if most of the power is at large scales, the gradients in $\ldi$ should be dominated by the largest allowed wavelengths. If $\ld \sim \ell_0$, this implies $\alpha \approx 1 / 2 \pi$. As discussed in section 4.2 below, our simulations show
$\alpha \simeq 0.17$, 
which is indeed close to $1/2\pi$. Note that while we do find a weak dependence of $\ldi$ on $\ma$, we are ignoring it in our analytic treatment, as including it does not lead to an improvement in the accuracy of our formula in the range of parameter space $( 0.7 < \ma < 5.0)$ that we have considered. 
We were unable to obtain converged results for higher equilibrium values of $\ma$
with our $512^3$ simulations, so we were unable to confirm the $1/\ma$ scaling found
by PZN00.

For small values of $\radlo$, AD will smooth the field fluctuations, 
so that $\ldb>\ldi$. Hence, equation (\ref{eq:vda4}) provides an upper limit on
$\vda$ if $\ldi$ replaces $\ldb$:
\beq
\frac{\vda}{\vrms^2}<\frac{2\ma^2}{3\alpha^2\radld^2 (1 + \frac{2}{3}\ma^2)}=
\frac{2}{3\alpha^2\ma^2 (1 + \frac{2}{3}\ma^2) \tau^2}.
\label{eq:vda1}
\eeq

This equation provides a good estimate of $\vda$ at high values of $\radld$, but 
it is too large at moderate or low values of $\radld$.
First, $\vda$ must be less than $\vrms^2$; in fact, in the weak coupling limit, $\vda=\frac 23\ \vrms^2$.
In Paper II, we introduced several different regimes
of AD: Regime 1 is ideal MHD, Regime 2 is standard AD ($t_f>t_{ni}\gg t_{in}$),
Regime 3 is strong AD ($t_{ni}>t_f>t_{in}$), Regime 4 is weak coupling
($t_{in}>t_f$) and Regime 5 is the hydrodynamic limit ($t_f/t_{in}\rightarrow 0$ for fixed
$\ma$). In Paper II, we found that
most molecular clouds are in Regime 2.
Here we shall consider regimes 1-3, omitting regime 4,
in which the heavy-ion approximation breaks down, and regime 5 in which it is irrelevant.
In regimes 1-3, the ions are reasonably well-coupled parallel to the field, so that AD
is primarily normal to the field. We therefore require 
\beq
\vda\leq \frac 23\,\vrms^2,
\label{eq:vda2}
\eeq
where the inequality approaches equality in the 
strong AD limit.
We also require that
the AD dissipation rate be less than the total dissipation
rate  ($\avg{\Gad}<\avg{\Gamma_t}$), so that
\beq
\avg{v_d^2}<\frac{\epsilon_t}{\tau}\;\vrms^2.
\eeq
In principle,we could include the factor $\frac 23$ here also, but it does not improve the accuracy of the
fit to the simulations. 

\subsection{THE AD HEATING RATE}

We have now found three upper limits on $\vda$; one of them is a good estimate of
the value of $\vda$ at large values of $\radld$ (Equation (\ref{eq:vda1})), and one 
is a good estimate at small values of $\radld$ (Equation(\ref{eq:vda2})). Combining these two
estimates and ensuring that $\vda$ is
less than all three upper limits, we adopt
\beq
\frac{\vrms^2}{\vda}\simeq\frac 32+\frac{\tau}{\epsilon_t}+\frac 32 \alpha^2\ma^2 \left(1 + \frac{2}{3}\ma^2\right) \tau^2,
\eeq
so that
\beq
\avg{\Gad}\simeq\frac{\tau}{\dis \frac 32+\frac{\tau}{\epsilon_t}+
\frac 32 \alpha^2\ma^2 \left(1 + \frac{2}{3}\ma^2\right) \tau^2}\left(\frac{\bar\rho\vrms^3}{\ld}\right).
\label{eq:avggad1}
\eeq

We now consider the values that the normalized AD heating rate,
$\avg{\Gad}/(\rho\vrms^3/\ld)$, 
takes in various limits. 
Varying one quantity at a time, we find that
the normalized AD heating rate approaches $\epsilon_t$ for $\vrms\rightarrow 0$;
0 for $\vrms\rightarrow\infty$; 0 for $\brms\rightarrow 0$;
$\epsilon_t\tau/(\frac 32\epsilon_t+\tau)$ for $\brms\rightarrow\infty$; and
0 for both $\bar\rho\rightarrow 0$ and $\bar\rho\rightarrow\infty$, 
holding the fractional ionization constant in both cases.

The careful reader will recall the comment that equation (\ref{eq:avggad}) is not
very accurate at large $\radld$. In order to overcome this problem, we determine
$\alpha$ by fitting our final result (Eq. (\ref{eq:avggad1}) or (\ref{eq:avggad2}))
to our simulation results. Fitting the results at the largest value of $\radlo$
that is well converged ($\radlo=113$), we find 
\beq
\alpha\simeq0.17.
\eeq
Our ideal MHD models at $\beta=0.01,\,0.1$, and 1.0 show that $\alpha$ is almost independent of $\ma$. 

\begin{figure}
\begin{center}
\epsscale{.80}
\includegraphics[scale=0.4]{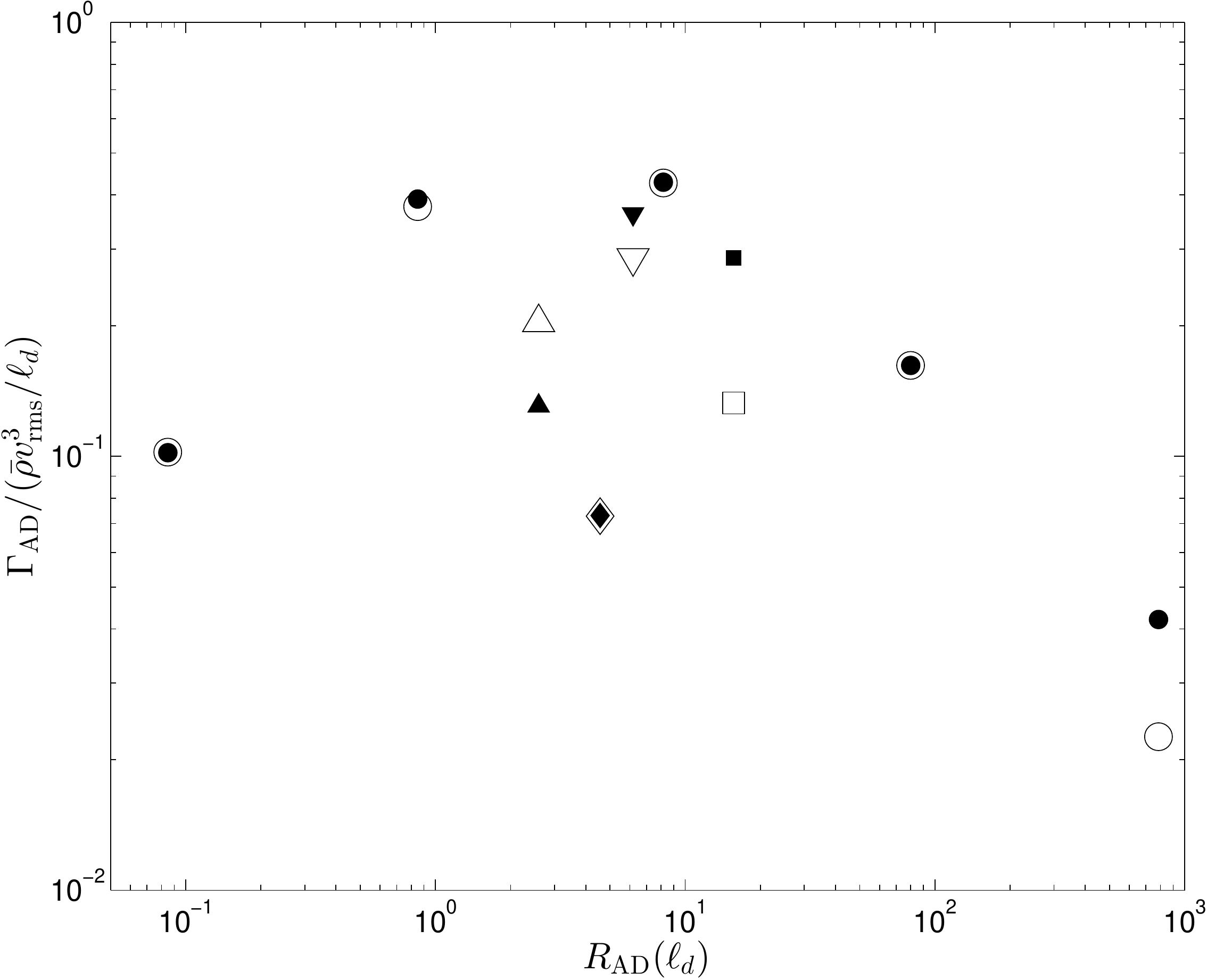}
\caption{Normalized AD heating rates of (a) five AD models (m3c2r-1 to m3c2r3) with initial $\mrms = 3$ and $\beta_0 = 0.1$ (solid circles), (b) Model m3c2r1b0 with initial $\mrms = 3$ and $\beta_0 = 1.0$ (solid down triangle), (c) Model m3c2r1b1 with initial $\mrms = 3$ and $\beta_0 = 10$ (solid diamond), (d) Model m10c2 with initial $\mrms = 10$ and $\beta_0 = 0.1$ (solid up triangle), and (e) Model m3c2r2b1 with initial $\mrms = 3$ and $\beta_0 = 10$ (solid square).  The open symbols are the predicted values from Equation (\ref{eq:avggad1}), using $\epsilon_t=0.55$ and the rms $\ma$ at equilibrium.  All predictions are within a factor of two of AD heating rates measured from simulations.
\label{fig4}}
\end{center}
\end{figure}

The AD heating rate in Equation (\ref{eq:avggad1}) can then be rewritten as
\beq
\avg{\Gad}\simeq\frac{1}{\dis1+\epsilon_t\left[\frac{3\ma^2}{2\radld}+0.043\radld \left(1 + \frac{2}{3}\ma^2\right) \right]}\left(\epsilon_t\,\frac{\bar\rho\vrms^3}{\ld}\right),
\label{eq:avggad2}
\eeq
which explicitly shows that it is always less than the turbulent dissipation rate.
The value of $\epsilon_t$ is shown in Figure \ref{fig2}; it has a value $\simeq 0.65$
for low values of $\radlo$ (including the hydrodynamic case), rises to a maximum
$\simeq 0.84$ at $\radlo\simeq 10$ and then falls to about 0.3 for large values
of $\radlo$ (including the ideal MHD limit).
We plot the normalized AD heating rates measured from our AD simulations versus $\radld$
in Figure \ref{fig4} together with the predicted AD heating rates from Equation (\ref{eq:avggad1}) using $\epsilon_t = 0.55$, the mean value from all our models listed in Table 1.
The predicted AD heating rates are all within a factor of two of the simulation values.

Numerically, the total turbulent dissipate rate (Equation \ref{eq:gt}) is
\beq
\Gamma_t=\epsilon_t\,\frac{\bar\rho\vrms^3}{\ld} = 3.8\times 10^{-28}
\left(\frac{\epsilon_t}{0.5}\right)\frac{\nbh v_5^3}{\ldpc}~~~~~\mbox{erg cm\eee\ s\e},
\label{eq:gt2}
\eeq
where $\nbh$ is the density of H nuclei, $v_5=\vrms/(10^5$~cm~s\e), and $\ldpc
=\ld/(1$~pc).
In estimating the heating rates in observed clouds,
one can replace $\ld$ with the observed cloud size, $\ell_{\rm obs}$.
Gas in molecular clouds typically has $\ma\sim 1$ \citep{cru99} and 
$\radlo\sim 20$ \citep{mck10}.
Figure \ref{fig2} shows that $\epsilon_t\simeq 
0.5$ for this value of $\radlo$. For these parameters, the AD heating rate is only slightly lower than the total dissipative heating: Equation (\ref{eq:avggad2}) implies
$\epsad/\epst\simeq 0.8$, whereas the numerical results for the model with the closest
set of parameters (m3c2r1, which has $\radolo=12$) gives $\epsad/\epst\simeq 0.7$.

\section{DISCUSSION}
How important is AD heating compared to cosmic-ray heating? Since AD heating
is a substantial fraction of turbulent dissipation for conditions that are typical in molecular
clouds, we compare with the latter. The turbulent line width is observed to increase with
scale as $\sigma=\vrms/\surd 3=\sigma_{\rm pc}(\ell_{\rm pc}/2)^{0.5}$, where typically
$\sigma_{\rm pc}=0.7$ km s$^{-1}$
\citep{mck07}. The turbulent heating rate in molecular clouds is then
\beq
\Gamma_t=2.4\times 10^{-28}\left(\frac{\epsilon_t}{0.5}\right)\left(\frac{\sigma_{\rm pc}}{0.7 \hspace{3 pt} \rm{km} \hspace{3 pt}  \rm{s}^{-1}}\right)^3\bar \nh
\ell_{\rm d\,pc}^{0.5}~~~~~\mbox{erg cm\eee\ s\e}.
\eeq
By comparison, the cosmic-ray ionization rate in molecular clouds with column densities $N_{\rm H}\la 2\times 10^{22}$~cm\ee, which has recently been revised upwards, is about $3.5\times 10^{-16}$~s\e\ (H$_2$ molecule)\e\ \citep{ind12}. \citet{gla12} find that the heating per ionization in molecular clouds is 13 eV,
so this ionization rate corresponds
to a heating rate of $\Gamma_{\rm CR}\simeq 3.6\times 10^{-27}\bar\nh$~erg~cm\eee~s\e. Under normal circumstances then, turbulent dissipation, including AD heating, is not competitive with cosmic-ray heating
in molecular clouds. There are significant variations in both the linewidth-size relation and the cosmic-ray
heating rate, however, so turbulent dissipation can dominate in some regions.
For very large molecular clouds ($\ell_d\ga 100$~pc), the two heating rates become comparable, but
it should be borne in mind that the heating due to turbulent dissipation is spatially localized: A single
shock wave at a velocity $\vrms$ that extends across the area of the cloud, $A$, 
could account for the bulk of the heating due to turbulent dissipation in the cloud, since the volume-averaged
heating rate is $\frac 12 \rho\vrms^3 A/V=\frac 12\rho\vrms^3/\ell\simeq\Gamma_t$, where $V$ is the volume
of the cloud and $\ell$ its size.

Finally, we compare our results on the AD heating rate in turbulent systems to those of PZN00, who examined this question previously. Note that PZN00 originally contained numerical errors that caused them to overestimate the mean AD heating rate. 
Therefore, we compare against their corrected numbers in PZN12 (when it is not important to distinguish between the two papers, we shall simply refer to PZN). We focus on their runs A1-A7, which they used to determine an expression for the mean AD heating rate. For convenience, we have summarized the key physical parameters of these runs in Table 2. 

\begin{table}
\begin{center}
\caption{Parameters of AD simulations A1-A7 in PZN \label{tbl-2}}
\begin{tabular}{cccccccc}
\\
\tableline\tableline
 Name & ${\cal M}_A$ & $\avg{|B|}$  & ${\cal M}^{~a}$  & $\vrms$  &  $\radlo$  & $\Gamma_{\rm{PZN}}$ & $\Gad^{~b}$\\ 
                    &     & $(\mu$G)     &                     & (km s$^{-1})$ &              & (erg cm$^{-3}$ s$^{-1}$) & (erg cm$^{-3}$ s$^{-1}$) \\
\tableline
A1	& 83.1     &   0.3         &   9.9     & 1.9  &   4.3e5  & 7.5e-30 & 2.9e-33 \\ 
A2	& 18.2     &   1.6         &  12.0     & 2.3  &   1.7e4 & 2.7e-28 & 2.7e-30\\ 
A3	& 8.2      &   2.6         &   9.1       & 1.7 &   4.6e3  &  4.3e-28 & 2.1e-29\\ 
A4	& 5.5      &   4.5         &  10.5      &  2.0&   1.8e3  &  1.8e-27 & 1.8e-28\\ 
A5	& 4.2      &   6.0         &  10.8       & 2.0 &   1.0e3  &  3.2e-27& 5.8e-28\\ 
A6	& 2.5      &  7.5         &   8.1       & 1.5 &   4.8e2    &   2.7e-27 & 1.3e-27 \\ 
A7	& 0.7      &  44.1         &  12.4      & 2.3 &   2.4e1  &   2.6e-25& 1.5e-25\\
\hline
\end{tabular}
\end{center}
$^a$Computed for 10 K gas.\\
$^b$From Equation (31).
\end{table}

PZN characterized the strength of AD in their simulations through the parameter $a$, which is related to our $\radlo$ by
\beq
\label{eq:atorad}
\radlo = \frac{\calm_A^2}{\calm} \hspace{3 pt} \frac{N}{a},
\eeq 
where $N$ is their numerical resolution. All the runs in Table 2 have $N=128$ and $a=0.21$, but the range in $\radlo$ is very large. 
Note, however, that PZN have no runs with $\radlo \lesssim 24$, so we cannot compare to them in that regime. We can, however, check for consistency at higher $\radlo$. 

From their simulations, PZN12 infer the mean AD heating rate in a turbulent molecular cloud 
for the case in which $a=0.21$:
\beq
\langle \Gad \rangle = 3.5 \cdot 10^{-26} \times \left( \frac{\langle |B| \rangle}{10 \,\mu {\rm G}} \right)^4 \left( \frac{\ma}{5} \right)^2 \left( \frac{\nbh}{520 \, {\rm cm}^{-3}} \right)^{-3/2} {\rm erg} \hspace{3 pt} {\rm cm}^{-3} \hspace{3 pt} {\rm s}^{-1}.
\label{eq:padadheat}
\eeq
Observe that we have expressed the density in terms of the number of hydrogen nuclei, rather than the number of neutral particles, as they did. PZN00 estimate that $\Gad\propto a^{0.6}$ over the range $0.11<a<0.74$, but this is questionable since
$a$ depends on the numerical resolution, whereas the AD heating rate does not.

To compare their results with ours, we compute the mean heating rate twice for each run listed in Table 2, once using their result (Equation \ref{eq:padadheat}), which we label $\Gamma_{\rm{PZN}}$, and once using our Equation (\ref{eq:avggad2}),
which we simply label $\Gad$. 
The results are listed in Table 2. We find that the agreement is fairly good 
(within a factor $\sim 5$)
for $\ma\la 4$ and $\radlo\leq 1000$, 
which is the regime we explored directly. However, for their runs at high $\ma$ and high $\radlo$, the disagreement is much worse, and by run A1, we are lower by about 3 orders of magnitude. 

What accounts for the difference? While PZN find that the magnetic length scale is proportional to $1 / \ma$, we find it to be only weakly dependent on $\ma$ in the range we were able to consider $(\ma < 5.0)$, and treat it as constant in our analytic theory. Although we could improve our agreement with PZN by inserting a $1 / \ma$ dependence (actually, we would need a $1/(1+\ma)$ dependence to be consistent with our results at low $\ma$) into the third term in the denominator in Equation (\ref{eq:avggad2}), we have chosen not to do so, since our simulation results do not provide evidence for this dependence, at least over the limited
range of $\ma$ we could explore.  Note that run A1 has $\radlo = 4.3\times 10^5$, meaning that this run was very close to the ideal MHD limit. The AD heating rate in this regime is only a negligible fraction of the total turbulent heating. The regime in which AD heating is significant corresponds to the last few runs in Table 2, which is where our agreement is fairly good, especially considering the differences in our numerical approaches---e.g., we use a two-fluid treatment to their one, we assume ion conservation and they that the ionization is determined by cosmic rays, and our turbulence driving used a fixed velocity pattern whereas theirs was random in time and was applied as a source term in the momentum equation. Finally, we note that real molecular clouds are observed to lie mainly in the moderate $\radlo$ regime \cite{mck10}. Thus, the agreement is reasonably
good in the regimes of the most physical and astrophysical interest.

\section{CONCLUSIONS}

We have discussed a number of the effects of ambipolar diffusion (AD) on weakly magnetized molecular clouds in Papers II and III, based on high resolution, 
two-fluid MHD turbulence simulations using the heavy-ion approximation. 
The strength of AD can be measured with the AD Reynolds number,
$\radlo$ (Equation \ref{eq:radell}).  
In the limit of low $\radlo$, one recovers the hydrodynamic limit, while in
the limit of high $\radlo$ one recovers ideal MHD. Molecular clouds
are observed to have intermediate values of $\radlo$, ranging
from 3 to about 70 in Crutcher's (1999) sample of molecular clouds with measured
field strengths \citep{mck10}.
In this paper, we focus on the heating due to the friction between ions and neutrals inside 
weakly ionized, turbulent molecular clouds over a wide range of conditions. 
We compute the AD heating rates directly from our two-fluid model using Equation (\ref{eq:adheat}). Our conclusions are:

\begin{itemize}
\item[1.] We find that the AD heating rate is a significant fraction of the overall turbulent heating rate in the range of $\radlo = 1 \sim 100$,
provided $\ma$ is not large. As noted above, this range of $\radlo$ encompasses
observed molecular clouds; furthermore,
observed molecular clouds typically have $\ma\sim 1$ \citep{cru99}.
Our AD turbulence simulations show that 
in this regime, up to 70-80\% of the total dissipation is in the form of AD heating.
AD heating gives a moderate increase in the total turbulent dissipation rate in molecular clouds with typical values of $\radlo$.  At smaller values ($\radlo \la 1$), the AD heating rate falls off rapidly as a result of infrequent collisions between ions and neutrals.

\item[2.] Heating due to turbulent dissipation, including AD heating, is generally less than cosmic-ray heating in molecular clouds, although there is substantial scatter in both rates.

\item[3.] AD significantly affects the length scale at which turbulent energy is dissipated. When AD is weak, either because the two fluids are too weakly coupled to cause significant momentum exchange ($\radlo\la 1$)
or because they are so well-coupled that no drift develops ($\radlo\ga 100$)
(Figure \ref{fig2}), almost all of the energy can cascade down to the viscous and/or resistive scales. On the other hand, when AD heating is strong ($\radlo \sim 1 - 100$, the regime most significant for molecular clouds), we find that most of the energy in the cascade can be removed by ion-neutral drift, with 
$\la \frac 12$ of the turbulent energy making it down to small scales. The scale at which turbulent energy is converted to heat can affect the temperature distribution of interstellar gas \citep{pan09}; however, we defer discussion of this effect to future work. 

\item[4.] We derive a relation (Equation \ref{eq:avggad2}) 
for the AD heating rate based on the global physical properties of the system.  The predicted ratio of the AD heating rate to the total is accurate to within a factor of two for all our AD models.  The relation is useful in predicting the AD heating rate inside molecular clouds from their observed properties.

\end{itemize}
\noindent

\acknowledgments
We wish to acknowledge Ellen Zweibel and Paolo Padoan for very helpful discussions.  Support for this research was provided by NASA through NASA ATP grant NNG06-GH96G (CFM, and PSL) and the NSF through grant AST-0908553 (CFM and ATM).  This research was also supported by grants of high performance computing resources from the National Center of Supercomputing Application through grant TG-MCA00N020.

\end{document}